\documentclass[prl,aps,amsfonts,amssymb,twocolumn,floats,showpacs]{revtex4}
\usepackage{graphicx}

\begin{document}

\title{Quantum Walks on a Random Environment}

\author{Yue Yin, D.E. Katsanos
and S.N. Evangelou\footnote{e-mail:sevagel@cc.uoi.gr}}
\affiliation{Department of Physics, University of Ioannina, Ioannina
45110, Greece}

\begin{abstract}

  Quantum walks are considered in a one-dimensional
  random medium characterized by static or dynamic disorder.
  Quantum interference for static disorder can lead to
  Anderson localization which completely hinders the
  quantum walk and it is contrasted with the decoherence effect
  of dynamic disorder having strength $W$, where a quantum to
  classical crossover at time $t_{c}\propto W^{-2}$ transforms
  the quantum walk into an ordinary random walk with diffusive spreading.
  We demonstrate these localization and decoherence
  phenomena in quantum carpets of the observed time
  evolution and examine in detail a dimer lattice
  which corresponds to a single
  qubit subject to randomness.

\end{abstract}

\pacs {03.67.Lx,72.15.Rn,03.65.Yz}

\maketitle

\par
\medskip
\section{I. INTRODUCTION}

\par
\medskip
During last decade quantum algorithms were proposed, such as
Grover's search\cite{r1}  and Shor's factorization\cite{r2}, which
can in principle perform certain computational tasks
quantum-mechanically, much more efficiently than their classical
counterparts. The related idea of quantum walks was also
introduced\cite{r3,r4,r5,r6} which generalize the classical random
walks widely used in various computations as the basis of
classical algorithms. The quantum walks are similar to classical
random walks
 but with a "quantum coin" operation which
replaces the coin-flip randomness in between each moving step on a
lattice or a graph. The state of the quantum coin which uniquely
determines the subsequent movement direction can also exist in
quantum superpositions, something impossible in the classical domain
where the coin has a specific outcome. In analogy with classical
random walks the quantum walks are expected to be useful for
designing quantum algorithms. For example, Grover's algorithm can be
combined with quantum walks in a quantum algorithm for "glued trees"
which provides even an exponential speed up over classical
methods\cite{r7}.

\par
\medskip
The main advantage of quantum walks is a highly improved behavior
over their classical counterparts since quantum wave propagation
is superior than classical diffusion. For example, the ballistic
mean square variance $\sigma^{2}(t)\propto t^{2}$ in the quantum
case can be compared to the linear spread law
$\sigma^{2}(t)\propto t$ of classical diffusion. This quadratic
speed-up is a general feature of quantum search
algorithms\cite{r1} and is also familiar from standard quantum
evolution of tight-binding electron waves on a periodic
lattice\cite{r8}. In quantum walks the classical probabilities
$P(x,t)$ are replaced by complex probability amplitudes
$\Psi(x,t)$ computed from the unitary dynamics of the
Schr$\ddot{o}$dinger's equation. The corresponding probability
amplitudes are determined by summing up over all possible paths of
propagation. Furthermore, to describe wave propagation in lattices
or graphs one does not need a "quantum coin"\cite{r9} and a
related continuous-time versions of quantum walks have been
introduced\cite{r10}. The discrete and continuous-time quantum
walks have recently been related to Dirac and
Schr$\ddot{o}$dinger's equations, respectively\cite{r11}.

\par
\medskip
We consider a continuous-time quantum walk via the equivalent
problem of a quantum particle initially localized in
one-dimensional lattice in the presence of static or dynamic
disorder. For a tight-binding electron at an integer lattice site
labeled by $x$ in one dimension  at time $t$ the wave function
$\Psi(x,t)$ obeys the linear wave equation ($\hbar=1$)
\begin{equation}
   \imath  \partial {\Psi(x,t)}/\partial{t} =
    \epsilon(x,t) \Psi(x,t)+\Psi(x-1,t)+\Psi(x+1,t),
\end{equation}
with  $\epsilon(x,t)$ an $x$-dependent random variable for static
disorder which is also $t$-dependent for dynamic disorder, where
lengths are measured in units of the lattice spacing and energies
or inverse times in units of the hopping integral. In order to
study quantum walks via Eq. (1) we choose the initial condition of
a particle at the origin $x=0$ with $\Psi(x,t=0)= \delta_{x,0}$
and characterize the quantum motion by the second moment for its
position
   $\sigma^{2}(t)  = \sum_{x}|x|^{2} P(x,t)$,
where $P(x,t)=|\Psi(x,t)|^{2}$ is the probability density. In the
absence of disorder ($\epsilon(x,t)=0$) the amplitude is given by
the Bessel function so that $P(x,t)= (J_{x}(2 t))^{2}$, where the
order $x$ of the Bessel function measures the distance travelled
from the origin while its argument is proportional to time $t$.
The evolution of the wave-packet at time $t$ was shown\cite{r8} to
display two sharp ballistic fronts at $x=\pm2t$. From properties
of the Bessel functions inside the spatial region $[-2t,2t]$ the
probability density $P(x,t)$ is an oscillating function multiplied
by $1/t$ while outside this region, denoted by the two ballistic
peaks, $P(x,t)$ decays exponentially. The ballistic mean-square
displacement is $\sigma^{2}(t) = 2 t^{2}$.

\par
\medskip
In the absence of disorder the quantum evolution of wave-packets
which occurs via the evolution operator $\exp(-iHt)$ with
Hamiltonian $H$ is very different from classical diffusion where
any initial state converges to a Gaussian steady state. An
initially squeezed $\delta$-function spatial wave-packet has
reduced spatial uncertainty and behaves like a quantum particle
consisting of all the eigenstates of $H$. Alternatively, a
spatially uniformly distributed initial wave-packet
$\Psi(x,0)=1/\sqrt{N}$, for every $x$ in an $N$-site chain,
consists of few eigenstates of $H$ near the lower band edge only.
Since the latter choice emphasizes states from the band edge the
semi-classical asymptotics is relevant\cite{r12,r13,r14}. The
wave-packets which initially consist of many eigenstates can be
related via the uncertainty principle to ultrashort laser pulses
of femtosecond duration.  The evolution of such coherent
superpositions of quantum states is realized in the physics of
trapped atoms in optical lattices\cite{r15}, trapped
ions\cite{r16}, etc.

\par
\medskip
Classical random walks in perfect one-dimensional lattices are
defined by the probabilities $p_{x}=q_{x}=1/2$ which determine the
random left and right motion from site $x$. This externally
induced randomization leads to classical diffusion for long $t$
with a Gaussian $P(x,t)$ and $\sigma^{2}(t) = 2D  t$, $D$ the
diffusion coefficient. For random walks in random media the
probabilities $p_{x},q_{x}=1-p_{x}$ become random variables
themselves, for example, they could be chosen from a flat
distribution within $(0,1)$. This is the so-called random random
or Sinai walk\cite{r17,r18} which leads to ultra slow classical
evolution $\sigma^{2}(t)\propto \ln t^{4}$, very close to a
complete cease.

\par
\medskip
The problem addressed in this paper concerns the fate of quantum
walks in a random environment, with both static and dynamic
disorder. To answer this question for static disorder we shall
combine previous knowledge from the field of wave propagation in
the presence of randomness where the quantum phenomenon of
Anderson localization\cite{r19} takes place (for its consequences
for quantum walks see\cite{r20}). We shall show that static
disorder is responsible for exponentially suppressed quantum
evolution with variance $\sigma^{2}(t)$ reaching a
time-independent limit for long $t$, depending on the strength of
static disorder and space dimensionality. Surprisingly, classical
random walks for static disorder are still propagating, although
ultraslowly\cite{r17,r18}. Other generalizations of discrete-time
quantum walks in aperiodic or fractal media by using biased
quantum coins have given slower (sub-ballistic) quantum
evolution\cite{r21}.  For dynamic disorder by coupling the quantum
system to a random environment decoherence occurs\cite{r22}  and
quantum physics becomes classical so that a quantum walk is still
propagating but only diffusively.

\par
\medskip
The main reason for examining the robustness of quantum walks in
the presence of noise is because disorder is unavoidable in most
quantum systems. Static disorder also appears for electrons in
lattices with permanent modifications due to impurities. The
dynamic disorder in this case could be driven by time-dependent
vibrations of the lattice atoms which have an impact on the
electronic site-energies. Apart from describing such
electron-phonon interactions, dynamic disorder also addresses the
presence of time-dependent noise in the memory qubits of quantum
computers. In this paper we demonstrate that although static
disorder hinders the motion of quantum walks due to negative
quantum interference from Anderson localization  via multiple
scattering from impurities, instead, for a dynamically random
environment the time-dependent disorder acts as a decoherence
mechanism at a crossover time $t_{c}$ which randomizes the quantum
walks and turns the quantum motion into classical. The discrete
finite space chosen in the simulations could be also used on a
finite computer. This general scheme for discretizing space is not
only suggested by solid state applications but it is, somehow,
related to the discreteness of the quantum information itself.

\par
\medskip
The paper is organized as follows: In Sec. I we introduced the
reader to the subject of quantum communication by setting the aims
of our quantum walk simulation in random media. In Sec. II we
briefly review the properties  of quantum walks in ballistic and
disordered one-dimensional media by showing quantum carpets which
demonstrate the difference between static and dynamic disorder.
Static disorder is shown to be responsible for negative quantum
interference of Anderson localization which stops completely the
quantum walk while dynamic disorder permits only diffusive
evolution of classical random walks. In Sec. III we display  the
quantum to classical crossover for dynamic disorder and consider a
qubit subject to dynamic disorder. Finally, in Sec. IV we
summarize our main conclusions.

\par
\medskip
\section{ II. QUANTUM CARPETS}

\par
\medskip
We have created space-time $x-t$ structures for the probability
density $P(x,t)$ on a one-dimensional finite $N$-site orthonormal
lattice space without disorder, with $t$-independent static
disorder and also rapidly varying dynamic disorder. The white
color in the figures denotes high probability density and the
darker colors lower values. In Fig. 1 a state is initially
released in the middle of the chain and the probability density
$P(x,t)$ is obtained by solving Eq. (1). In Fig. 2 the same is
done for a spatially uniform initial state.  For static disorder
Eq. (1) could be alternatively solved by considering the
time-evolution of a state vector $|\Psi(0)\rangle$ expressing the
probability density via the stationary eigen-solutions
$H|j\rangle=E_{j}|j\rangle$ with space-time wave function
\begin{equation}
   \Psi(x,t)  =  \sum_{j=1}^{N}e^{-iE_{j}t} \psi_{j}(x)
   \langle j|\Psi(0)\rangle ,
\end{equation}
with the amplitude on site $x$ denoted  by $\psi_{j}(x) = \langle
x|j\rangle$. For time-dependent disorder Eq. (1) was solved via a
fourth-order Runge-Kutta algorithm.

\par
\medskip
In Fig. 1 we present our results for the initial choice of a
$\delta$-function in the middle of the chain with
$|\Psi(0)\rangle=|0\rangle$ and in  Fig. 2 for a uniform initial
state $|\Psi(0)\rangle=(1/\sqrt{N}) \sum_{x=1}^{N}|x\rangle$ is
computed on finite chains where the wave-packet scatters from
their hard ends. In the rest of Figs. 3-8 our results are obtained
for a self-expanding chain with a $\delta$-function initial choice
to make sure that the wave-packet does not reach the boundaries.
This allows to study the quantum to classical crossover by
computing the mean-square-variance and the autocorrelation
function vs. time.

\par
\medskip
\begin{figure}
\includegraphics[angle=270, width=8.0cm]{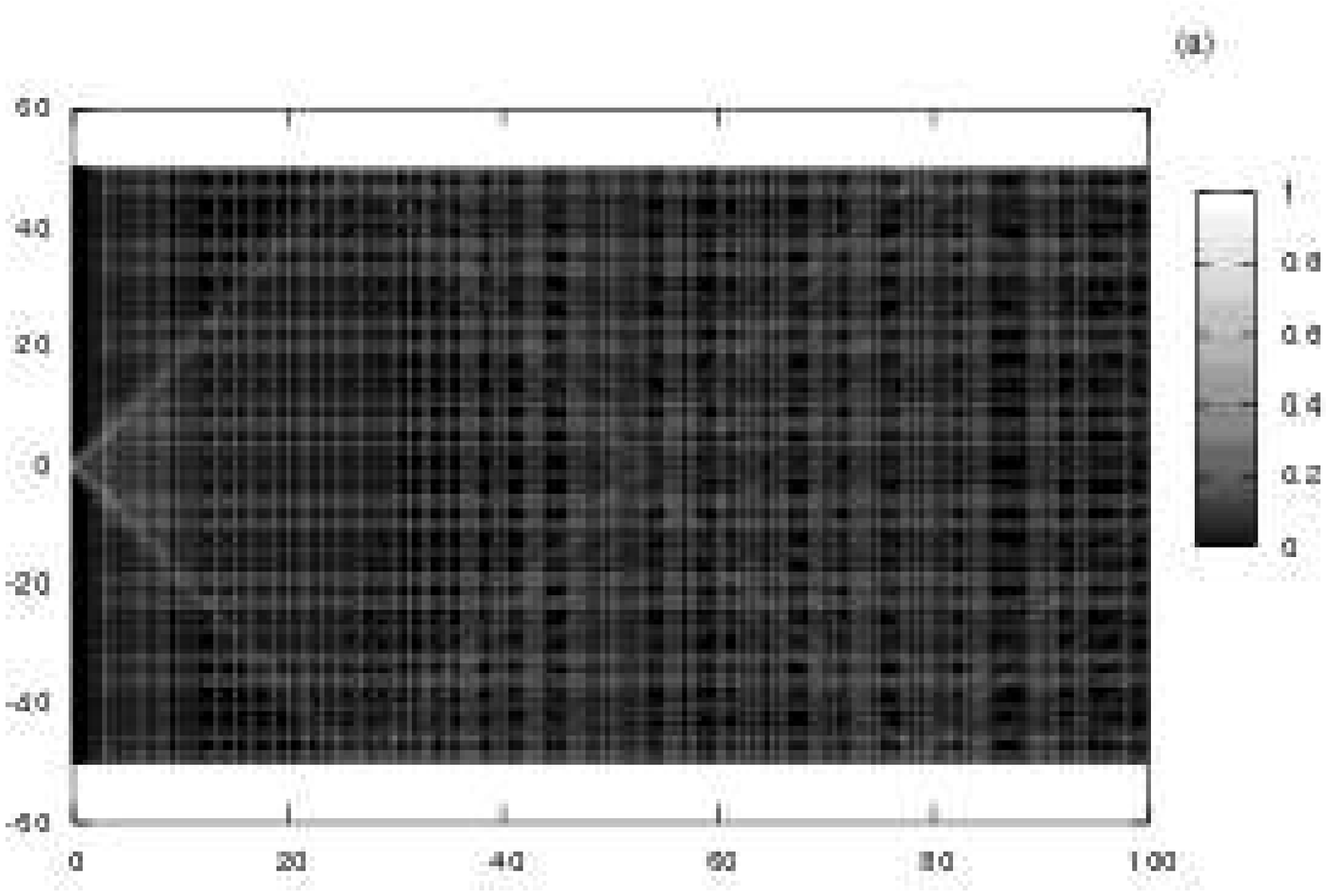}
\includegraphics[angle=270, width=8.0cm]{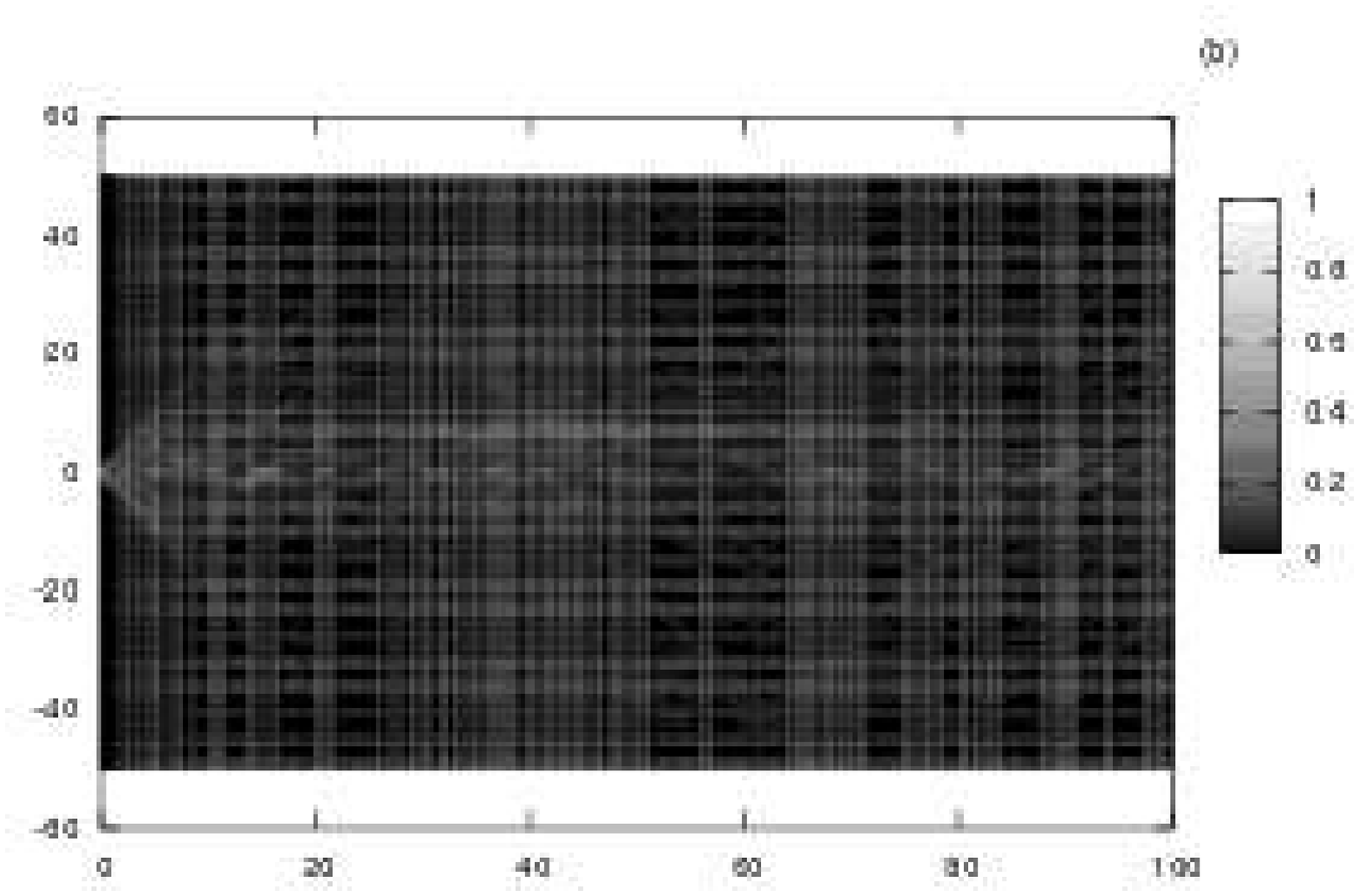}
\includegraphics[angle=270, width=8.0cm]{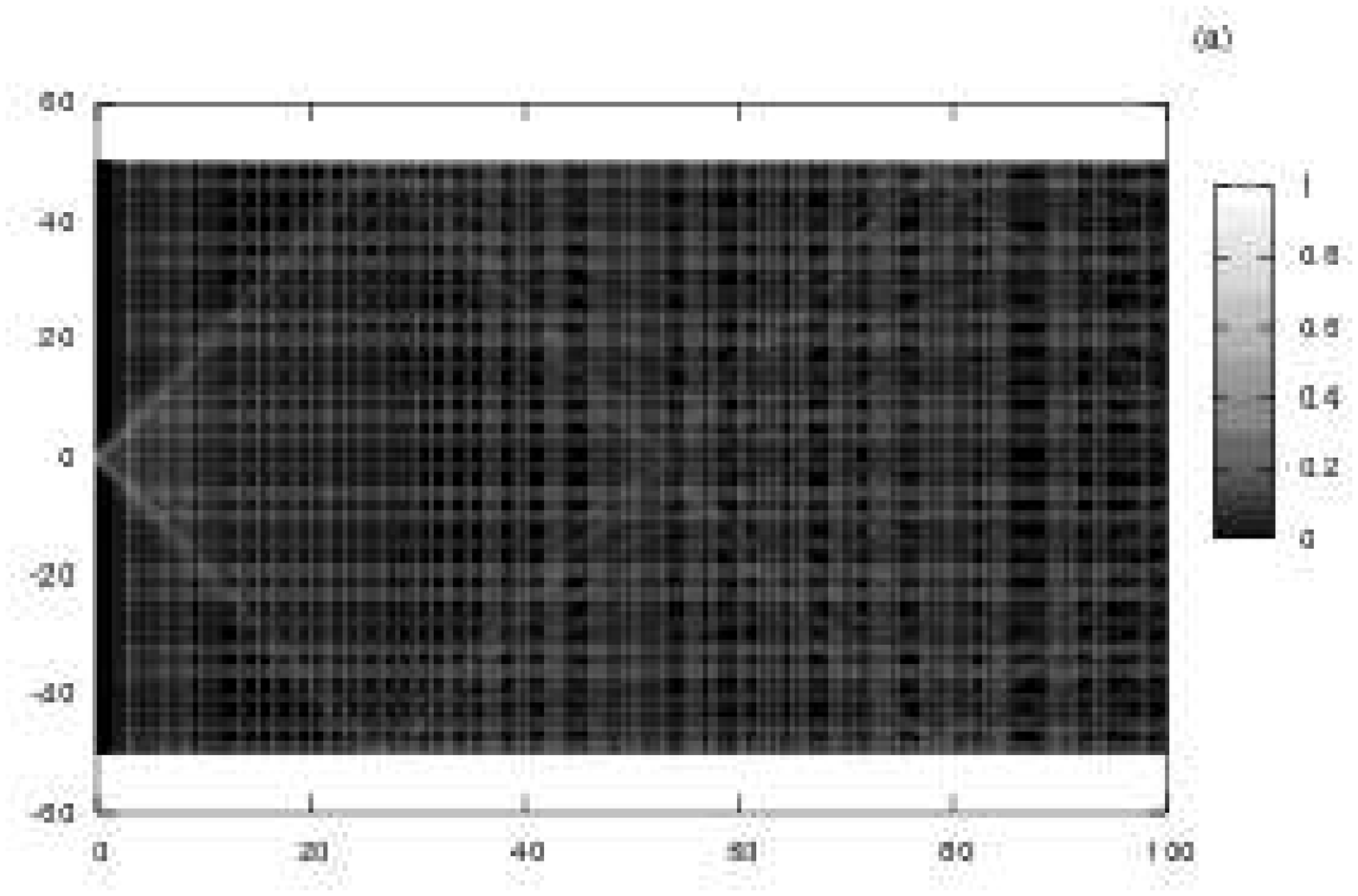}
\includegraphics[angle=270, width=8.0cm]{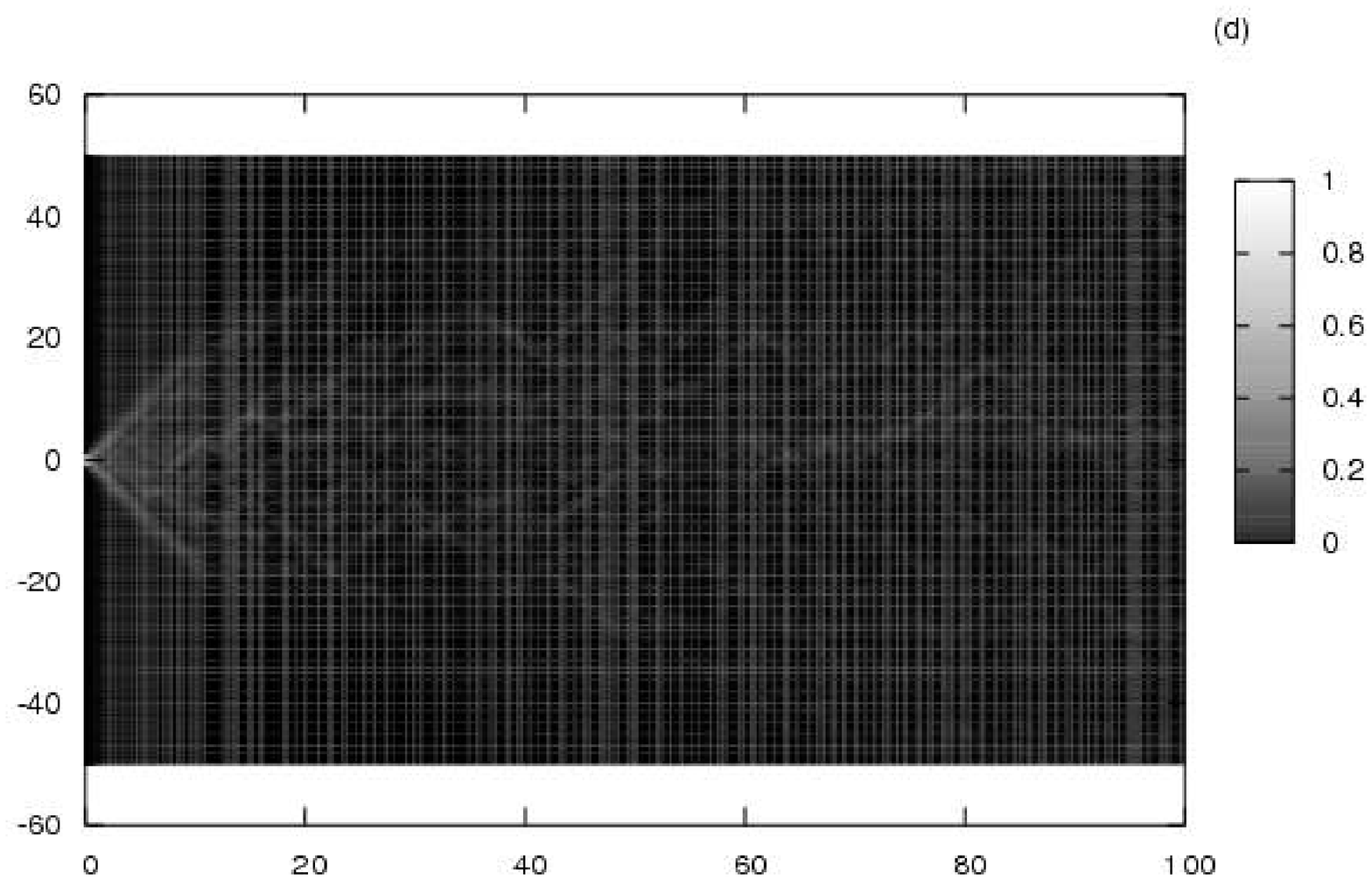}
\caption{Quantum carpets which show the probability density
$P(x,t)$, for $x$ in the vertical axis and $t$ in the horizontal
axis, are generated by an {\bf initial $\delta$-type} spatial
state for $t=0$ $\Psi(x,0)=\delta_{x,0}$ in the middle (left of
the figure) with chain length $N=101$. {\bf(a)} The ballistic case
for the absence of disorder where perfect quantum revivals can be
clearly seen. {\bf(b)} For static disorder of strength $W=1.5$
quantum interference causes Anderson localization which stops the
quantum motion and the probability to stay in the initial site
remains high. For stronger static disorder this probability
becomes even higher. {\bf(c,d)} For dynamic disorder with values
$W=1$ and $W=5$, one can see quantum interference only initially
for small $t$ on the left hand side of the figure where the
$\delta$-type wave-packet moves ballistically. After certain time
(for $W=1$ is estimated $t_{c}\approx 60$ and for $W=5$ about
$t_{c}\approx 3$) the quantum interference is lost while the
particle still moves but classically.}
\end{figure}

\par
\medskip
\begin{figure}
\includegraphics[angle=270, width=8.0cm]{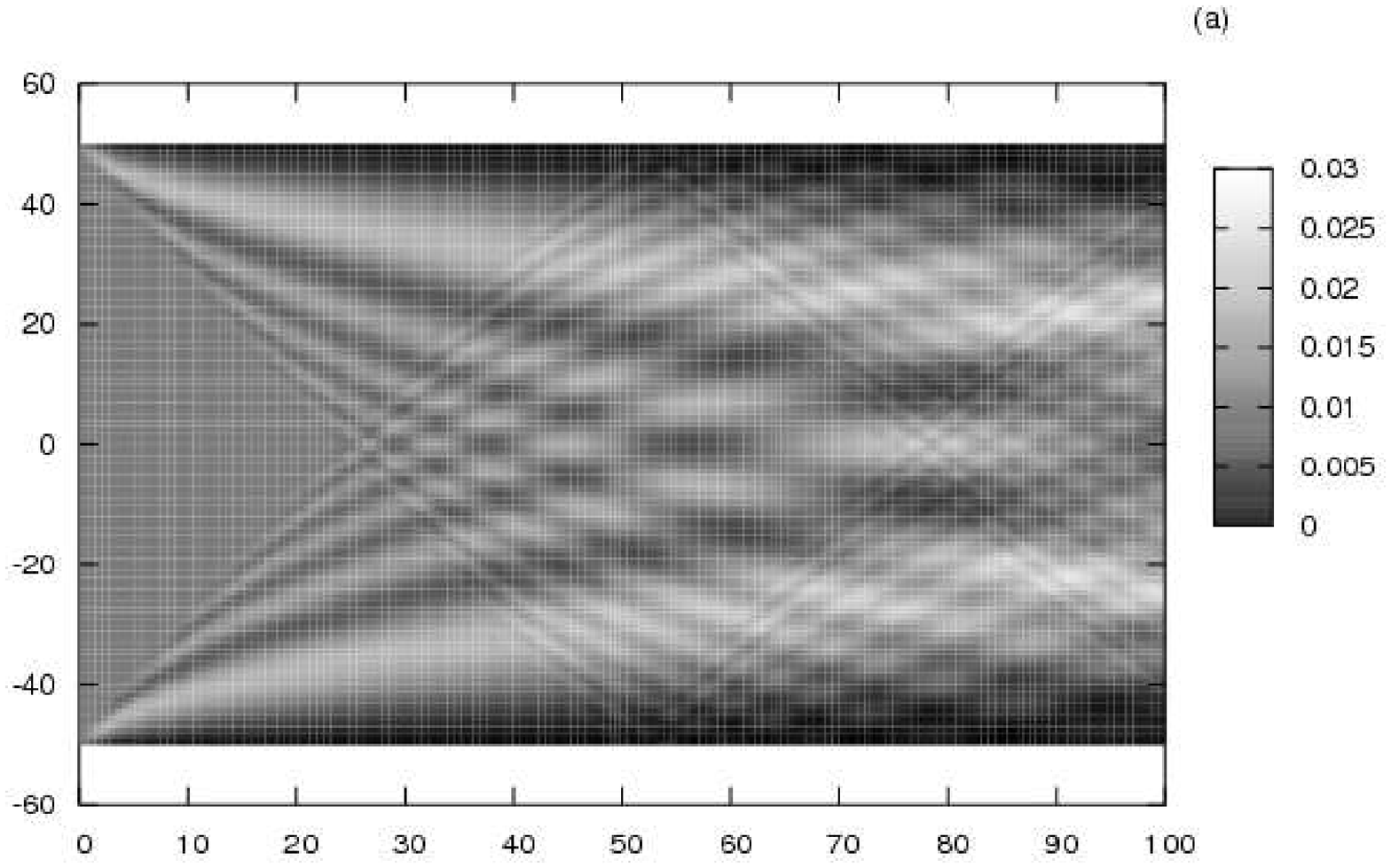}
\includegraphics[angle=270, width=8.0cm]{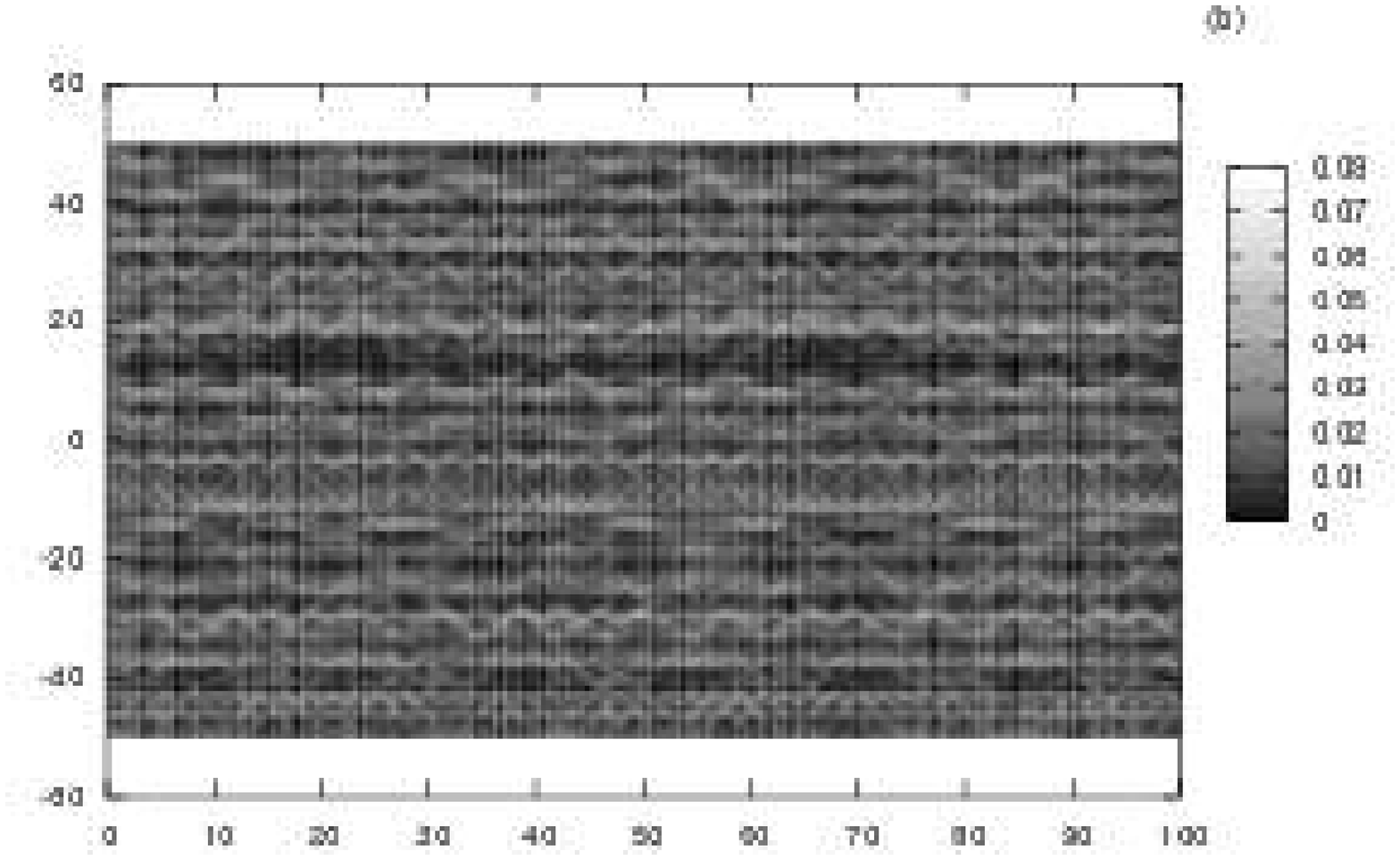}
\includegraphics[angle=270, width=8.0cm]{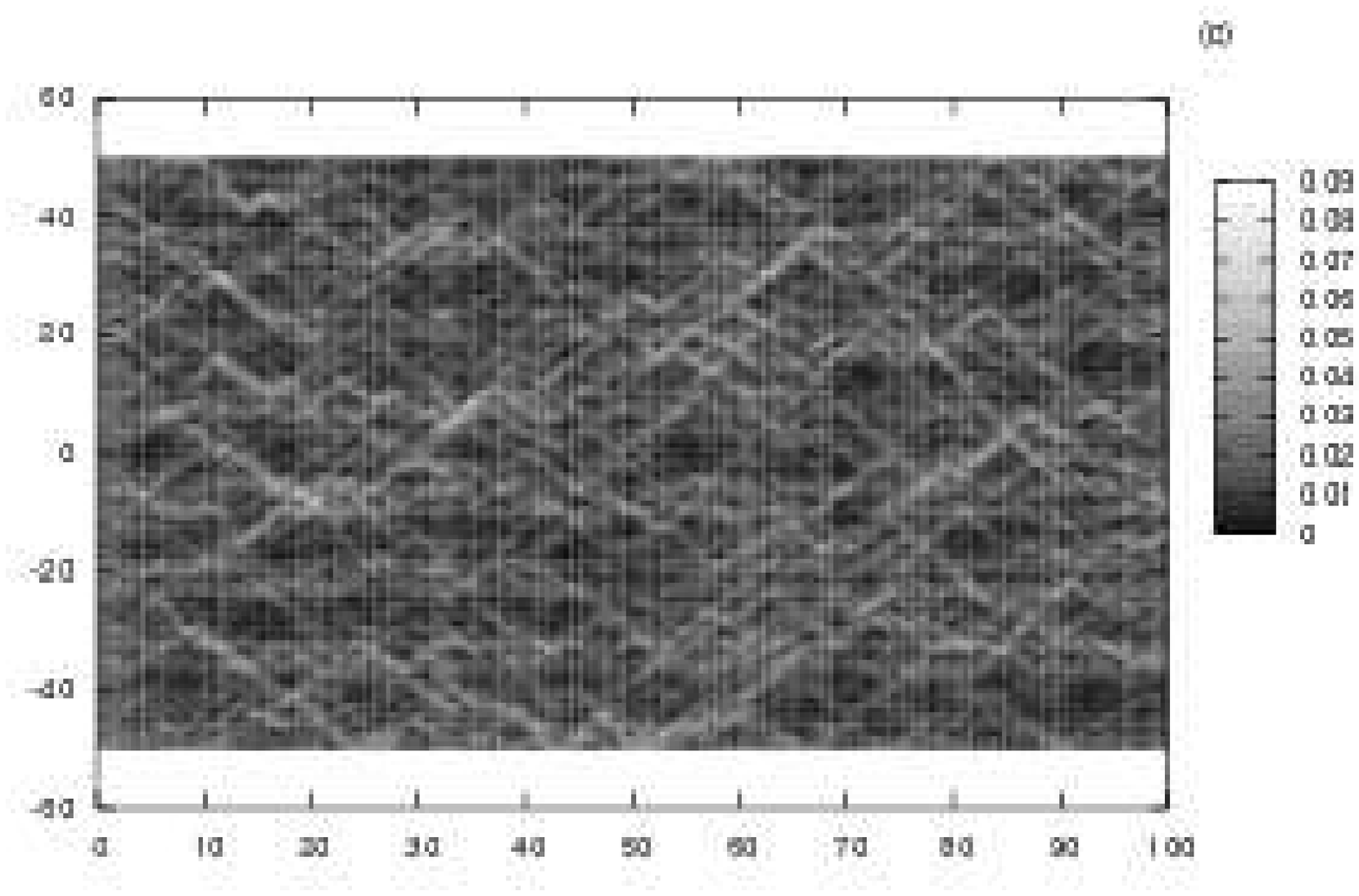}
\caption{Quantum carpets showing the probability density $P(x,t)$
for a {\bf spatially uniform} initial state $\Psi(x,0)=1/\sqrt{N}$
on a chain of length $N=101$. {\bf(a)} For the ballistic case in
the absence of disorder one can see recurrences for short times
which disappear for longer times where quantum interference
effects become apparent. {\bf(b)} For  static disorder of strength
$W=5$ Anderson localization occurs with localization length
$\xi\sim 100W^{-2}\sim 4$ much less than the system size $N=101$.
The displayed regions of high amplitudes show the positions where
the particle localizes. {\bf(c)} In the presence of dynamic
disorder $W=5$ the quantum interference effects vanish. The main
difference between dynamic disorder (c) and static disorder (b) is
that in (c) the regions with high values of $P(x,t)$ keep changing
leading to randomization so that the particle can still move but
in a "classical" fashion.}
\end{figure}

\par
\medskip
\subsection {ballistic motion}

\par
\medskip
In this case, obviously, quantum walks perform at their best. The
ballistic description is valid for solid state systems in the
absence of disorder which refers to the motion of a point particle
in an $N$-site chain with $\epsilon=0$ in Eq. (1) which gives
$E_{j}=2\cos({\frac{\pi j}{N+1}})$ and $ \psi_{j}(x) =\sqrt{{\frac
{2}{N+1}}} \sin({{\frac {{j\pi}{x}}{N+1}}})$, $j=1,2,\ldots,N$.
The corresponding space-time pictures are shown in Figs. 1(a) and
2(a). Quantum revivals can be seen where the particle returns to
its initial position and reconstructs like a classical particle
which moves with constant velocity reflecting at the boundaries of
the chain\cite{r14}. We observe that the quantum revivals of Fig.
1(a) and Fig. 2(a) do not repeat indefinitely but become less and
less accurately as time progresses. This is due to effects from
boundary scattering which become more prominent for broad
wave-packets in the right hand-side of Fig. 2(a) where noisy
evolution is established. The obtained fractal pattern  is a
result of peculiar quantum interference effects due to scattering
from the hard walls at the ends of the chain\cite{r14}.

\par
\medskip
\subsection {static disorder}

\par
\medskip
It can have dramatic consequences for quantum walks, particularly
in low dimensions, since for static disorder they can stop
completely due to Anderson localization. From Figs. 1(b) and 2(b)
we can see how strong static disorder with $\epsilon(x)$ chosen
from a uniform probability distribution within $[-W/2,+W/2]$
causes destructive interference with Anderson localization. This
crossover from ballistic motion to localization  has dramatic
consequences for quantum walks in one-dimension in the presence of
static disorder. In higher dimensions the Anderson transition from
extended to localized states is expected, via an intermediate
chaotic regime which is rather better for quantum propagation. The
probability density of Fig. 1(b) is shown to stay around the
middle site where the initial wave-packet has maximum amplitude
and it remains there for longer times. For the uniform initial
state of Fig. 2(b) the larger amplitudes remain on many sites
indefinitely.

\par
\medskip
\subsection {dynamic disorder}

\par
\medskip
A quantum walk can operate in the presence of dynamic disorder but
only for short times  since for longer times its motion becomes
entirely classical, indistinguishable from an ordinary random
walk. The effect of dynamic disorder is equivalent to introducing
coin chaos which makes the quantum coherence disappear\cite{r22}.
In order to see this decoherence effect we have chosen a random
$\epsilon(x,t)$ rapidly varying with $t$ by:

(i) Updating at random the diagonal site energies $\epsilon(x,t)$
at a time length comparable and often much smaller to the time
step of the numerical method.

(ii) Varying the diagonal energies by
\begin{equation}
\epsilon(x,t)=amp* \cos(\omega_{x}t+\phi_{x})
\end{equation}
where $amp$,  $\omega$ and $\phi$ are the amplitude, frequency and
phase for the motion of levels, respectively. We chose to vary the
frequency $\omega$  at random uniformly within the interval
$[0,2\pi]$ and fixed the phase $\phi_{x}$ to zero.

\par
\medskip
From Figs. 1(c), (d) and 2(c), we can see the effect of dynamic
disorder for the two initial wave-packets, $\delta$- function and
broad, respectively.  In Fig. 1(c) the quantum motion seen on the
left hand side of the figure quickly disappears  and this also
happens in Fig. 1(d) where classical diffusive motion is seen to
arise. In Fig. 2(c) the randomization effects of dynamic disorder
become more obvious and could be contrasted with quantum
localization for static disorder (Fig. 2(b)). The
quantum-to-classical crossover takes place after a characteristic
time $t_{c}\propto W^{-2}$.

\par
\medskip
\section{III. QUANTUM TO CLASSICAL CROSSOVER FOR DYNAMIC DISORDER}

\par
\medskip
\subsection{decoherence in an $N$-site chain}
\begin{figure}
\includegraphics[angle=0,width=6.0cm]{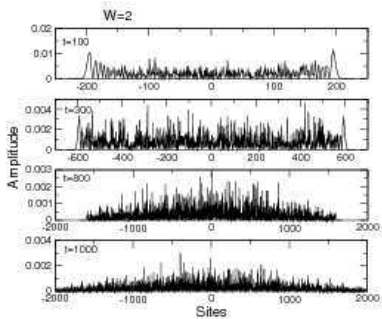}
\caption{ The probability density is shown to display the gradual
decoherence, from ballistic for small $t$ to diffusive evolution
for long $t$, with strength  of dynamic disorder $W=2$. The two
ballistic fronts in the  quantum case gradually disappear and the
shape approaches a Gaussian with classical diffusion.}
\end{figure}

\par
\medskip
The decoherence effect of dynamical disorder which turns the
quantum wave propagation into classical diffusion is shown in Fig.
3 for a self-expanding chain. The probability density $P(x,t)$
gradually changes from a shape displaying two ballistic peaks of
the quantum wave for small $t$ towards a Gaussian for large $t$.
In Fig. 4 the quantum to classical crossover is shown for the
mean-square-variance $\sigma^{2}(t)$ and the autocorrelation
function or return probability $C(t)={\frac
{1}{t}}\int_{0}^{t}P(0,t')dt'$. Eventually, the classical
asymptotic laws $\sigma^{2}(t)\propto t$ and $C(t)\propto
t^{-1/2}$ set in after an initial period of quantum ballistic
motion where $\sigma^{2}(t)\propto t^{2}$ and $C(t)\propto
t^{-1}$. In Fig. 5 the effect of sinusoidal dynamic disorder  is
considered with constant amplitude and randomly varying the phase.
The results are similar to Fig. 4 with the approach to the
classical limit even faster in this case. From Figs. 4 and 5,
except for no difference between the two types of dynamic
disorder, we find that the crossover region between the ballistic
law for small times and the diffusive law for long times is smooth
having a mixed quantum and classical character.

\begin{figure}
\includegraphics[angle=0,width=6.0cm]{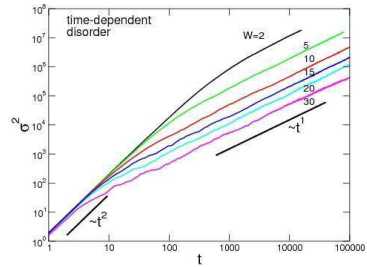}
\includegraphics[angle=0,width=6.0cm]{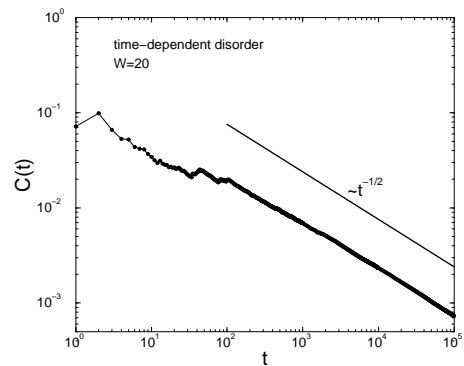}
\caption{{\bf (a)} Log-log plot of the mean-square -displacement
or variance $\sigma^{2}(t)$ vs. time $t$ for various values for
the dynamic disorder $W$. The crossover from ballistic quantum
motion to diffusive classical motion occurs at earlier times as
$W$ increases. {\bf (b)} The autocorrelation function or "return
to the origin" probability $C(t)$ vs. $t$ for $W=20$ displays
classical diffusive behavior.}
\end{figure}

\begin{figure}
\includegraphics[angle=0,width=6.0cm]{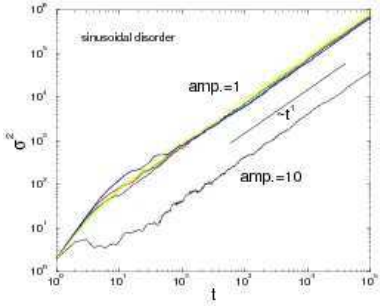}
\includegraphics[angle=0,width=6.0cm]{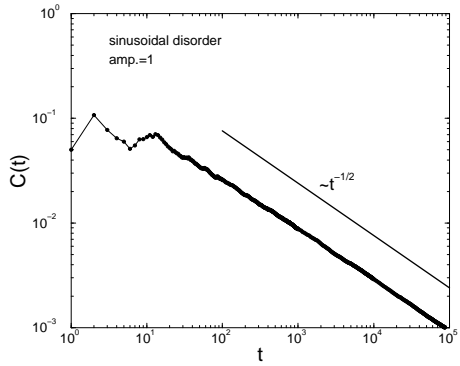}
\caption{  {\bf (a)} The $\sigma^{2}(t)$ vs. time $t$ same as in
Fig. 4(a) but for the sinusoidal dynamic disorder of Eq. (4). {\bf
(b)} The $C(t))$ vs. time $t$ same as in Fig. 4(b) but for the
sinusoidal dynamic disorder of Eq. (4).}
\end{figure}

\par
\medskip
The effect of dynamic disorder on the quantum evolution is
displayed in the linear and log plots of Fig. 6 which show the
snapshots of the evolving spatial wave for an initial
$\delta$-function in a self-expanding chain. The decoherence
effect of dynamical disorder is seen from the rapid approach to a
Gaussian shape. The complete phase diagram is summarized in Fig. 7
with $t_{c}$ vs the strength of dynamic disorder $W$ which
displays a wide crossover grey color region where the law
$t_{c}\propto W^{-2}$ is approximately obeyed.

\begin{figure}
\includegraphics[angle=0,width=6.0cm]{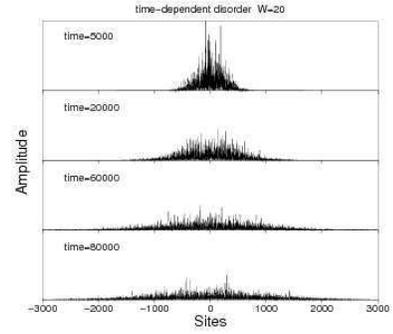}
\includegraphics[angle=0,width=6.0cm]{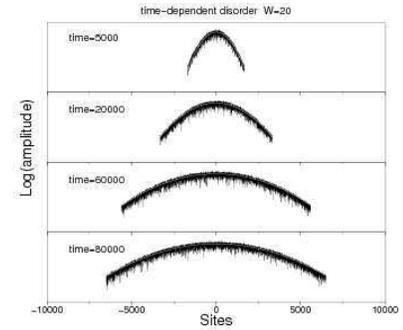}
\caption{The snapshots for dynamic disorder of strength W=20 in a
self-expanding chain where the approach to a Gaussian is seen. {\bf
(a)} The probability density $P(x,t)=|\Psi(x,t)|^{2}$ as a function
of space $x$ at fixed times $t=5000$, $t=20000$, $t=60000$ and
$t=80000$. {\bf (b)} The same as in (a) but for the log of the
amplitude.  }
\end{figure}

\begin{figure}
\includegraphics[angle=0,width=6.0cm]{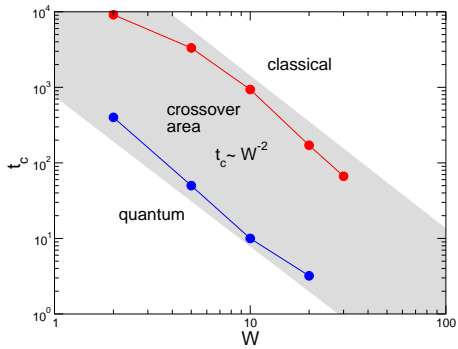}
\caption{ The quantum to classical crossover for dynamic disorder
of strength $W$ occurs at $t_{c}\propto W^{-2}$. This is shown by
the grey area which was estimated from two sets of points
connected via lines for the quadratic ballistic law to stop (blue
line) and the linear diffusive law to begin (red line),
respectively.}
\end{figure}

\par
\medskip
\subsection{decoherence in a qubit}

\par
\medskip
We have examined in detail the quantum walk in a random two-level
system ($N=2$) which has recently attracted attention in the
context of quantum information processing. The operation of qubit
and logical gates in the presence of a noisy environment is
important for understanding quantum computers. The usual noise for
such two-level system is usually due to various sources while
non-Gaussian randomness arises from hopping background charges for
different statistically independent fluctuators.

\par
\medskip
The time-dependent Hamiltonian is
\begin{equation}
 H= \left( \begin{array}{cc}
     \epsilon_{1}(t) & \gamma  \\
     \gamma & \epsilon_{2}(t) \\
  \end{array}
\right)
\end{equation}
with random diagonal terms defined by
\begin{equation}
     \langle \epsilon_{i}(t) \rangle =0,
      \langle \epsilon_{i}(t)\epsilon_{j}(t') \rangle =\delta
      \delta_{i,j}\delta(t-t'),i,j=1,2,
\end{equation}
where $\delta=W^{2}/12$ measures the disorder chosen from a box
distribution within $[-W/2,+W/2]$. The averaged matrix elements of
the density matrix $\bf{\rho}$ can be obtained from a decoupling
suggested in\cite{r23}
\begin{equation}
  \imath\overline{\dot{\rho_{11}}}=
-\imath\overline{\dot{\rho_{22}}}=\gamma
(\overline{\rho_{21}}-\overline{\rho_{12}})
\end{equation}
\begin{equation}
  \imath\overline{\dot{\rho_{12}}}=-2\imath \delta
\overline{\rho_{12}}+\gamma(\overline{\rho_{22}}-\overline{\rho_{11}}))
\end{equation}
and if we define
\begin{equation}
 \overline{\rho_{11}}={\frac {1}{2}} +\rho,
 \overline{\rho_{22}}={\frac {1}{2}} -\rho,
  \overline{\rho_{12}}=R+iJ,
  \overline{\rho_{21}}=R-iJ,
\end{equation}
the corresponding equations become
\begin{equation}
 \dot\rho=-2\gamma J,
 \dot R=-2\delta R,
 \dot J=-2\delta J + 2\gamma \rho.
\end{equation}
Their general solutions (with appropriate constants)
\begin{equation}
\left(
  \begin{array}{c}
    \rho \\
    J \\
  \end{array}
\right)
 =C_{+} \left(
  \begin{array}{c}
    -2\gamma \\
    \Lambda_{+} \\
  \end{array}
\right) e^{\Lambda_{+}t} +C_{-} \left(
  \begin{array}{c}
    -2\gamma \\
    \Lambda_{-} \\
  \end{array}
\right) e^{\Lambda_{-}t},
\end{equation}
\begin{equation}
R=C_{R} e^{-2\delta t},
\Lambda_{\pm}=-\delta\pm\sqrt{\delta^{2}-4\gamma^{2}}
\end{equation}
by choosing as initial state one of the two levels  with $
\rho(0)=1/2, R(0)=J(0)=0$.

\par
\medskip
Finally, the averaged off-diagonal matrix element of the density
matrix is easily shown to be
\begin{equation}
\overline{\rho_{12}(t)}=\imath J(t)= {\frac {\imath \gamma}
{\sqrt{4\gamma^{2}}-\delta^{2}-}
\sin(\sqrt{4\gamma^{2}-\delta^{2}}t)e^{-\delta t}}
\end{equation}
for $\delta^{2}<4\gamma^{2}$.  If $\delta^{2}>4\gamma^{2}$ in Eq.
(12) the quantity under the square root changes sign and $\sin$ is
replaced by $\sinh$. Thus, for $t\to \infty$ the averaged density
matrix approaches half the unit matrix with only diagonal matrix
elements and the quantum coherences described by
$\overline{\rho_{12}}$ becoming zero, oscillating for the quantum
case $\delta<2\gamma$ and monotonically for the classical case
$\delta>2\gamma$.

\par
\medskip
In order to examine the dephasing effect of dynamic disorder we
have plotted in Fig. 8 the phase $\theta$ of $ \rho_{1,2}$ vs. $t$
obtained from numerical computations.  Our results are presented
for $\gamma=1$ and different values of disorder $W$ which verify
the critical value $W_{c}=\sqrt{24}$ of the previous analysis
based on averages. The quantum coherence remains for weak dynamic
disorder $W< W_{c}$ while for higher dynamic disorder $W> W_{c}$
the phase randomizes and the system becomes classical.

\begin{figure}
\includegraphics[angle=0,width=6.0cm]{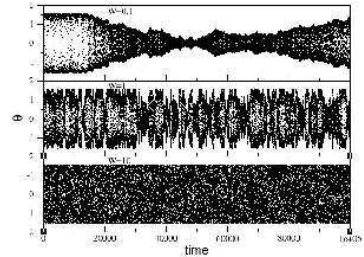}
\caption{The phase $\theta$ of the off-diagonal matrix element of
the density matrix $ \rho_{1,2}$ vs. time for a two-level system
with dynamic disorder $W=0.1$, $1$ and $10$. The decoherence due
to dephasing is seen for the highest value $W=10>W_{c}$ where
$\theta$ completely randomizes.}
\end{figure}

\par
\medskip
\section{IV. DISCUSSION}

\par
\medskip
Quantum walks are quantum analogues of classical random walks
which have been proposed for quantum computation purposes to
create quantum algorithms which run faster in quantum computers.
They can also arise from mapping various physical problems (e.g.
see \cite{r24}). Some quantum algorithms which speed-up classical
methods have already been successfully employed, such as for
search problems on graphs. These algorithms which show amplitude
amplification during the evolution could be efficiently
implemented in a quantum computer. However, since they are often
confronted with disorder we have examined  how quantum
wave-packets move in the presence of disorder, by computing the
probability density $P(x,t)=|\Psi(x,t)|^{2}$ from solving the
time-dependent Schr$\ddot{o}$dinger equation in the discrete space
$x$ of one-dimensional lattice. The static disorder due to
imperfections and the dynamic disorder due to the environment
could become obvious in scattering from nanostructures or can
appear from environmental noise, averaging over measurements, etc.

\par
\medskip
Our main conclusions from the quantum evolution of
$\delta$-function and very broad initial spatial wave-packets are:
(1) In random media quantum walks can perform even worse than
their classical counterparts since Anderson localization from
negative quantum interference completely stops the quantum walk
although the corresponding ordinary random walks in the presence
of disorder can still move despite infinitely slowly. (2) For
dynamic disorder we have no benefit from quantum walks either,
since for longer times the ballistic evolution for small-$t$
crosses over to classical diffusion for long-$t$ and the quantum
walks become classical via a quantum to classical crossover. (3)
The answer to the question "what slows down the quantum walk?" is,
on one hand, "static disorder via negative quantum interference"
and, on the other hand, "dynamic disorder at long enough times
which slows down the quantum walk and makes it no different from
ordinary random walk". Therefore, quantum interference in random
media can hold surprises for quantum walks and their advantages
should appear for weak disorder or short times only. In higher
dimensions quantum walks are also expected to operate for weak
disorder to avoid quantum localization. In conclusion, our
computations show Anderson localization or decoherence as the main
enemies of quantum walks in the presence of static and dynamic
disorder, respectively, which destroy their well-known quadratic
or exponential speed-up. Our study could be useful towards
creating better quantum search algorithms in the presence of
disorder\cite{r25}.


\end{document}